%% file: main.tex
\documentclass[journal]{IEEEtran}

\usepackage{cmap}
\usepackage[T1]{fontenc}
\DeclareFontShape{T1}{ptm}{m}{scit}{<->ssub * ptm/m/it}{}
\usepackage[utf8]{inputenc}
\usepackage{amsmath}
\usepackage{amssymb}

\usepackage{graphicx}
\graphicspath{{images/}}
\usepackage{booktabs}
\usepackage{array}
\usepackage{multirow}
\usepackage{longtable}
\usepackage{pbox}
\usepackage[caption=false,font=footnotesize]{subfig}
\usepackage[inline]{enumitem}
\usepackage{xurl}
\usepackage{enumitem}
\usepackage{url}
\usepackage[hidelinks,bookmarks=false]{hyperref}
\usepackage[nameinlink,noabbrev]{cleveref}

\begin{document}

\title{Sequential Semantic–Non-Semantic Watermarking: Robustness Against UnMarker Attacks}

\author{
Md Ifraham Iqbal\\
\small Department of Data Science,
Friedrich-Alexander-Universität Erlangen-Nürnberg (FAU), Erlangen, Germany\\
\small \href{mailto:ifraham.iqbal@gmail.com}{ifraham.iqbal@gmail.com}
\quad
\href{https://orcid.org/0000-0002-3628-045X}{ORCID: 0000-0002-3628-045X}
}

\maketitle

\begin{abstract}
\input{sections/abstract}
\end{abstract}


\input{sections/01_introduction}

\input{sections/03_methodology}

\input{sections/04_results}
\input{sections/05_discussion}

\input{sections/06_conclusions}

\bibliographystyle{IEEEtran}
\bibliography{references}

\section*{Acknowledgment}

The author thanks Dr. Daniel Tenbrinck for his supervision
and guidance during the master's thesis research on which this work is based.

The author also gratefully acknowledges Friedrich-Alexander-Universität
Erlangen-Nürnberg (FAU) for providing access to the high-performance
computing (HPC) resources used to conduct the experiments in this study.

\begin{IEEEbiography}
[{\includegraphics[width=1in,height=1.25in,clip,keepaspectratio]
{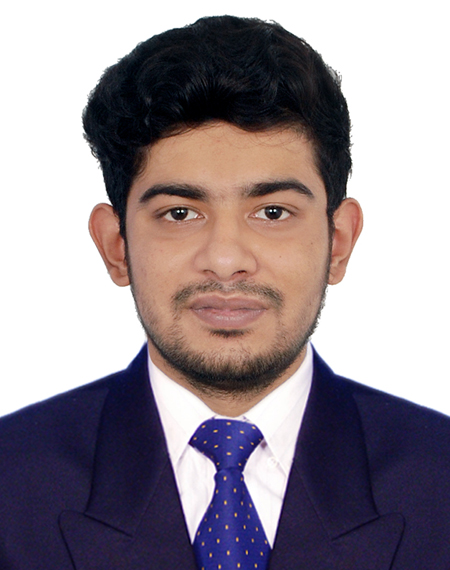}}]
{Md Ifraham Iqbal}
received the B.Sc. degree in Computer Science and Engineering from
United International University, Dhaka, Bangladesh, in 2020, and the
M.Sc. degree in Data Science from Friedrich-Alexander-Universität
Erlangen-Nürnberg (FAU), Erlangen, Germany, in 2026.

His research has covered machine learning, deep learning, computer
vision, health informatics, robotics, and tabular-data learning. His
current research interests include defensive image watermarking,
watermark-removal robustness, computer vision, and robust AI systems.
He has authored and coauthored research published in IEEE Access,
IEEE Region 10 Symposium (TENSYMP), and Springer proceedings, with
work spanning convolutional neural networks for tabular data, disease
prediction, intelligent transportation, and robotics.
\end{IEEEbiography}


\end{document}

%% file: sections/abstract.tex
Advancements in artificial intelligence (AI) have led to a rapid rise in digital content generation, thereby increasing risks of large-scale fraud and scams. The global counterfeit market is expected to cross 1.79 trillion dollars. Hence, regulations were made for the safe use of AI and the watermarking industry is projected to grow. Non-semantic and semantic watermarking schemes have been developed to tackle these digital scams. Alongside watermarking schemes, watermark removal attacks have advanced significantly. Recently, UnMarker introduced a non-semantic attack and a semantic attack that could remove any watermark at low computational cost with minimal visual fidelity loss. 

Motivated by this threat, the study introduces Sequential Semantic Non-Semantic watermarking (SSNSW) schemes that combine a semantic and non-semantic watermarking scheme. UnMarker has a key limitation that it targets either high-frequency or low-frequency watermarks, not both simultaneously. To address this weakness, the study uses SSNSW schemes as a layered defense mechanism. The results showed that when an SSNSW image is subjected to an UnMarker attack, one watermark remains highly decodable: StegaStamp under the non-semantic attack and HiDDeN under the semantic attack. This further illustrates that traceability and robustness can be improved effectively by combining two methods rather than excessively optimizing one method, which risks higher image deterioration.

Afterwards, the study tested sequential UnMarker attacks on the SSNSW schemes. Results show that sequential semantic and non-semantic attacks remain effective in removing dual watermarks at a higher perceptual cost. The findings suggest that two types of watermarks should be combined to boost robustness and force attackers to use higher perceptual costs to remove them. The study also indicates that layered watermarking and attack strategies are more effective than single-stage methods, thereby providing a concrete direction for future multi-layered watermarking and watermark removal pipelines.

%% file: sections/01_introduction.tex
\section{Introduction}
\subsection{Background}
Watermarking have historically been used for embedding identification marks into physical objects \cite{laurentius2016italian}. Later, it became crucial for paper currencies and tackling counterfeiting\cite{regula2024watermarks}. During the late 1900s, digital media emerged and with it rose key concerns of pirated products \cite{karaganis2011media}. Based on this necessity, the term 'digital watermarking' was introduced by Tirkel and Osborne \cite{van1994digital}. 

\begin{figure}[!t]
    \centering
    \includegraphics[width=1\linewidth]{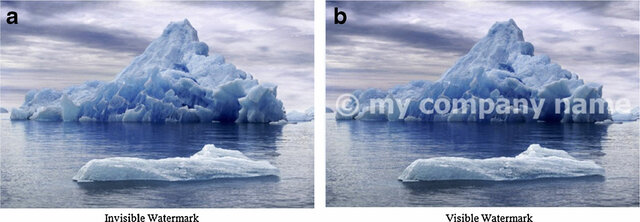}
    \caption{Illustration of visible and invisible watermarking. Source: adapted from \cite{youssef2014adaptive}.}
    \label{fig:visible_invisible}
\end{figure}

Digital watermarking refers to embedding invisible signatures into images that verify ownership and protect products from piracy. The difference between a visible and invisible watermark is shown in Figure \ref{fig:visible_invisible}.
Digital signature marks the source of the image while preserving quality and remaining imperceptible. Digital watermarking is used for authentication, provenance tracking, fingerprinting and device control.

The growth in technology and the emergence of generative Artificial Intelligence (genAI) have made piracy and fabrication of digital products extremely easy. The creation of AI-based content has become the core of many businesses \cite{malakar2024generative}. Group-IB \cite{huang2025groupib_voice_deepfake_scams} report shows that this will continue to rise and projects that by 2027 these AI-enabled losses will reach up to \$40 Billion from \$12 Billion in 2023\cite{huang2025groupib_voice_deepfake_scams}. 
Furthermore, the global counterfeit market is projected to reach a valuation of 1.79 trillion dollars by 2030 \cite{corsearch_counterfeit_trade_1_79t_2030}. Reports show that from the year 2022 to 2023, deepfake fraud cases increased by a staggering 1740\% as scammers adopted AI tools\cite{khan2024deepfakes}. 

In addition, in the first quarter of 2025, these AI-driven scams crossed a market value of \$200 Million \cite{colman2025wef_detecting_dangerous_ai}. These scams also include AI-enabled video call-based impersonation and have led to heists that are worth a high amount. An example in \cite{areghan2023data} shows a \$25 Million theft from a UK-based company using deepfakes. Financial Institutions report that each fraud incident creates average losses of \$600 thousand dollars  and multiple cases are worth over a million dollars \cite{financialpost_regula_deepfake_600k} worth of losses alone. The speed and realism make these deepfake scams highly effective at extorting huge amounts of money. Deloitte reports that financial firms saw deepfake-related fraud attempts surge 700\% in their contexts. \cite{lalchand2024deloitte_deepfake_banking_fraud}. Furthermore, unauthorized image reuse also directly affects creators and brands. An
estimated 2.5 billion images are stolen every day, resulting in losses
of approximately \$600 billion
\cite{Artaius2019ImageTheft25Billion}. AI has also made it easier to
reuse brand images, causing financial losses and reducing consumer
trust \cite{Meliana2025AIFailsBrandTrust}.

\begin{figure}[t]
    \centering
    \includegraphics[width=0.85\linewidth]{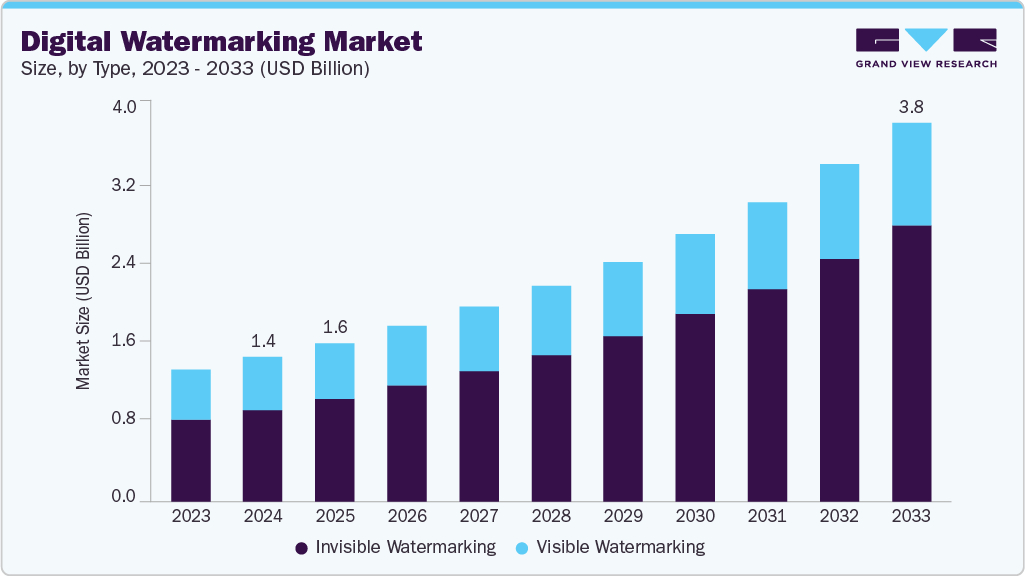}
    \caption{Global digital watermarking market size and projected growth. Source: adapted from \cite{GrandViewResearch2025DigitalWatermarkingMarketReport}}
    \label{fig:digital-watermarking-market-size}
\end{figure}

Due to these issues, major nations have issued regulations to make AI-generated content identifiable. The European Union introduced the Artificial Intelligence Act \cite{EuropeanParliament2025EUAIACTFirstRegulation} in 2023, which requires AI systems to ensure that AI-generated content is identifiable. Similarly, United States Executive Order 14110
\cite{USPresidencyProject2023EO14110SafeSecureTrustworthyAI} addressed the identification of AI-generated content. China's Deep Synthesis regulations, in effect since 2023, also require AI-altered or generated media to be identifiable \cite{Zhang2023ChinaDeepSynthesisProvisions}. China extended
these requirements in 2025 by requiring both visible and invisible labels for AI-generated media.

As a result, the digital watermarking market growth is expected to surge in the next few years. The digital watermarking market is valued at roughly \$1.4 to \$1.5 billion in 2024 \cite{GrandViewResearch2025DigitalWatermarkingMarketReport} and is expected to reach \$3 to \$4 billion in size by 2030 \cite{GrandViewResearch2025DigitalWatermarkingMarketReport} as industries invest in content tracking and anti-piracy. 63\% of this revenue is due to invisible watermarking schemes that are stealthy and do not leave a mark on images \cite{GrandViewResearch2025DigitalWatermarkingMarketReport}.

\begin{figure}[!t]
    \centering
    \includegraphics[width=1\linewidth]{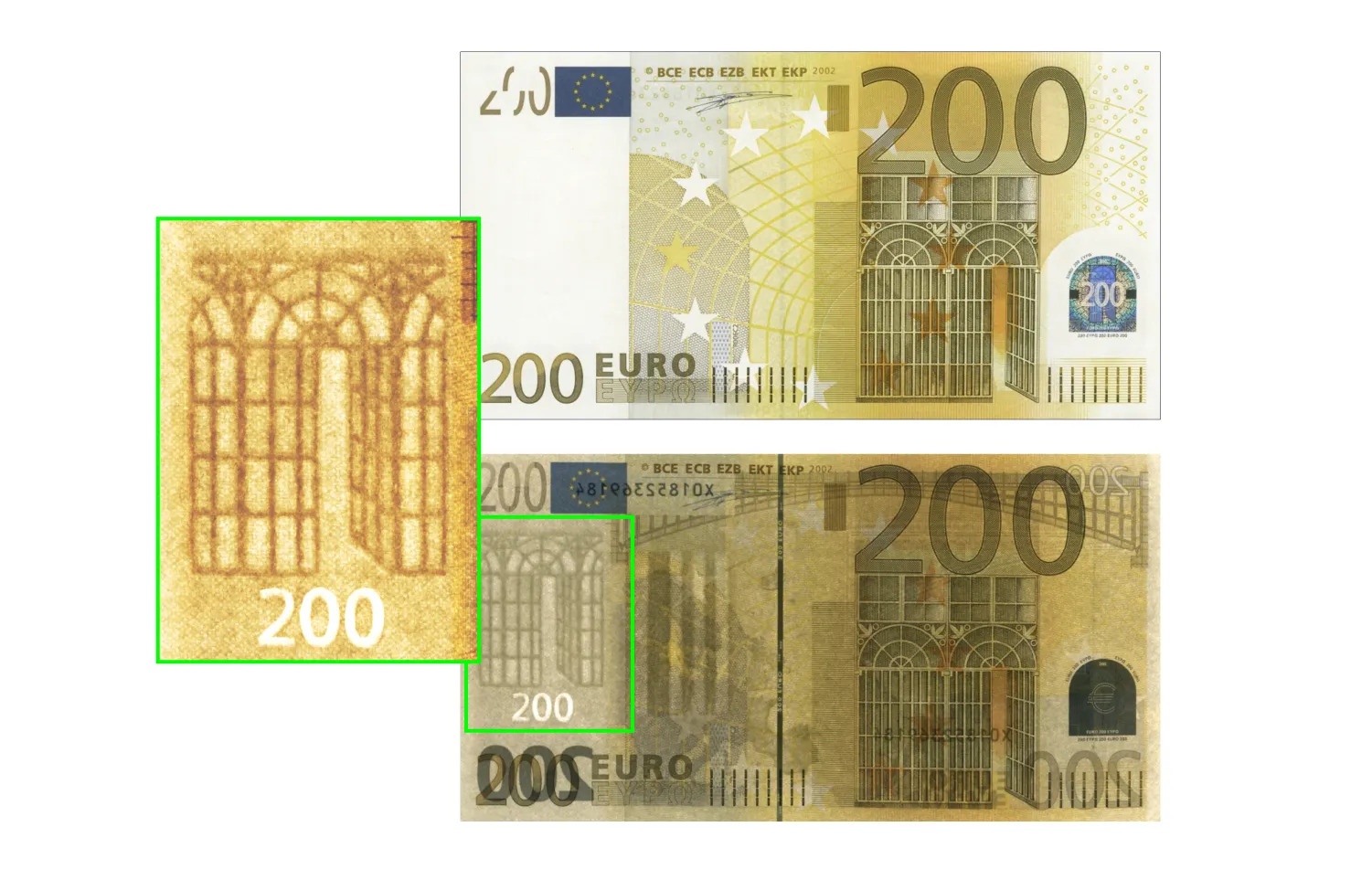}
    \caption{2002 Euro banknote watermark example. Source: \cite{regula2024watermarks}.}
    \label{fig:euro_watermark}
\end{figure}

\subsection{Previous Watermarking methods}

\paragraph{Traditional and Learning-Based Watermarking}
Traditional watermarking methods used spatial- and
frequency-domain techniques. Spatial domain approaches such as Least
Significant Bit \cite{zhu1999multiresolution} were computationally inexpensive but vulnerable to compression, noise, and image transformations.
Frequency-domain methods based on the Discrete Cosine Transform \cite{zhang2003efficient}
and wavelet transforms \cite{raval2003discrete} offered improved robustness. However, they
could still be disrupted by filtering, compression, and geometric
operations. Attackers who remove watermarks still found relatively simple ways to remove these watermarks using various noise or warping filters. Hence, interest grew in the use of Machine Learning to develop robust watermarking schemes.

These limitations motivated the adoption of
deep-learning-based watermarking, beginning with convolutional
approaches such as Hiding Images in Plain Sight
\cite{baluja2017hiding}. This was an early convolutional neural network (CNN) based approach to embed one image (watermark) inside another image (original). However, the embeddings was vulnerable to removal with JPEG compression. Authors of \cite{zhu2018hidden} built on the idea of HIPS to introduce HiDDeN, an end-to-end trainable model that embedded a message within any image while preserving perceptual similarity to original image. It consists of an encoder that embeds the message, a noise layer during training that simulated real-world distortions like cropping, blurring and compression, and a decoder that recovers the message. Additionally, an adversarial discriminator was used to encourage encoder to  produced images that looked visually similar with embedded watermarks. 

The paper showed that CNNs were capable of producing robust and invisible watermarks that were not easily removed with watermarking removal attacks. Building on the HiDDeN framework, ROMark\cite{wen2019romark} used adversarial techniques to improve robustmess against watermark removal attack, enhancing results. Similarly, RedMark\cite{ahmadi2020redmark} introduced a differentiable attack simulation layer to improve robustness against distortions (e.g., cropping, compression). This further improved the robustness of HiDDeN to cropping and compression attacks. Robust Invisible Video Watermarking GAN (RivaGAN )\cite{zhang2019robust} introduced deep-learning-based video watermarking using attention mechanisms during adversarial training. RivaGAN produced low-distortion watermarks that resisted video processing operations.

\paragraph{Physical and Generative Watermarking}
HiDDeN was designed to provide robustness against digital transformations and preserve image identity but, printing images and recapturing images rendered HiDDeN less effective.. They add further distortions like lighting changes, angles and required a more sophisticated noise layer. To address this issue, Tancik et al introduced StegaStamp \cite{tancik2020stegastamp}, a watermarking scheme for watermarking robustness in images after they are printed out and recaptured. The encoder-decoder concept of HiDDeN is used with a more comprehensive CNN architecture. Furthermore, an approximate differentiable noise layer is used to simluate lighting, camera focus focus, colour and other recapturing distortions. The study showed that upto 56 bit message could be embedded and then decoded using smartphone cameras. StegaStamp extended deep-learning-based watermarking robustness from digital to physical print scenarios

In recent years, generative AI (genAI) based models have started to produce realistic images. Hence, watermarking AI-based content has become necessary to reduce misuse. Ning Yu et al proposed the artificial fingerprinting method \cite{yu2021artificial}, which embeds model specific fingerprints during training rather than after image production. Fingerprints are added the training data so that generated images inherit a detectable signature
that can be recovered by a corresponding decoder.
The method was later extended using Responsible Disclosure of Generative Models \cite{yu2022responsible}, making fingerprinting more scalable and providing a path to fingerprinted AI digital products. Afterwards, Google developed their own SynthID Image \cite{gowal2025synthid} and SynthID Text \cite{dathathri2024scalable} that supports watermarking at an large industrial scale. They used imperceptible embeddings that enable the decoder to flag-AI generated content after image transformations. The emergence of SynthID enables deployment of genAI systems in the industry.

Retraining LLMs can be expensive and so Lukas et al, introduced Pivotal Tuning Watermarks (PTW) \cite{lukas2023ptw} to trains a small subset of model parameters so that generated outputs contain the embeddings. Existing image generates used PTW to train efficiently. The authors of Stable Signature\cite{fernandez2023stable} extended the idea to diffusion models. Stable Signature fine-tuned the decoder to producde embeddings while keeping the original architecture intact. A seperate watermark decoder is used to recover the embedding from the image which ensures that no post-processing is required. Stable Signature is highly robust and showed that it can detect watermarks with over 90\% accuracy even after 90\% of the image is cropped. Finally, Tree-Ring Watermarks (TRW) \cite{wen2023tree} adds the embedding in the initial noise vector of the diffusion model. The embedding resembles tree rings in \textbf{frequency space} and are passed through the denoising process of diffusion models and added to generated image. 
\begin{figure}[!t]
    \centering
    \includegraphics[width=1\linewidth]{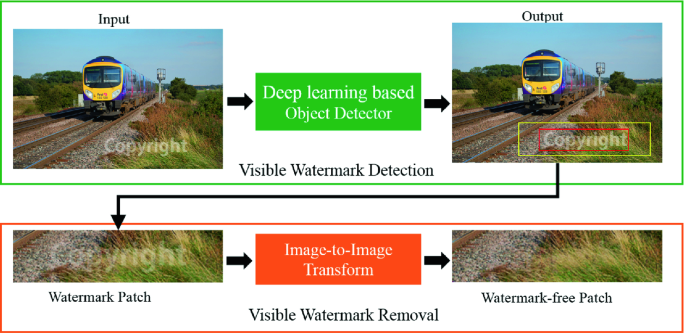}
    \caption{Example of visible watermark detection and removal. Source: \cite{cheng2018visible}.}
    \label{fig:watermark_removal}
\end{figure}



\subsection{Previous Watermark removal methods}
 
Watermark removal matured alongside the watermarking schemes. Early attacks relied on compression, geometric transformations and flitering. However, encoder-decoder based watermarking was highly robust and so attackers also started using deep learning methods to remove watermarks. An illustration of a visual watermark removal is shown in Figure \ref{fig:watermark_removal}. Dekel et al. \cite{dekel2017effectiveness} used multiple images with the same visible embedded watermark to be trained and then remove these watermarks. WDNet \cite{liu2021wdnet} split images into watermark and background components to remove them. Similarly, Split-Refine \cite{cun2021split} used pixel localization to  locate the visible watermarks and remove them. The Prior Attack study \cite{hu2024stable} uses information of the Stable Signature model to remove Stable Signature completely when given prior knowledge of the model. However, none of these methods were designed for invisible watermarks.

For invisible watermarks, WEvade \cite{jiang2023evading} randomized the decoder's embedded message. so that decoder accuracy falls to 50\%. However, the model required access to the targeted watermarking scheme. Then, Saberi et al \cite{saberi2023robustness} proposed an attack that introduces Gaussian noise and then passes image through a pretrained diffusion denoiser, enabling black box removal with no prior knowledge. The attack had a high success rate against non-semantic, subtle watermarks (e.g., HiDDeN). However, against semantic (low-frequency) watermarks, the method blurred parts of the image to remove these watermarks. CtrlRegen\cite{liu2024image} uses Control Network guided regenteration to preserve original image while destroying the watermark structure. This method is extremely effective but computationally expensive. 

Overall, these methods were limited by prior knowledge, model
access, computational cost, image degradation, or their inability to
address both semantic and non-semantic watermarks.UnMarker \cite{kassis2025unmarker} introduced two attacks to address these limitations. The authors argue that all watermark carriers are the spectral amplitudes. They introduced the non-semantic (high-frequency) and and semantic (low frequency) attack  to disrupt these frequencies that carry the watermarks.

The Non-semantic attack used Direct Fourier Loss to maximize spectral differences between the original watermarked and attacked image. This results in high frequency disruption with perceptual constrain bounds to limit visual changes. Low frequency regions are visible and so  thesemantic attack used learnable filters to disrupt low frequency regions as pixel manipulation will cause visual artifacts. Perceptual constraints are applied again. Results showed the UnMarker substantially reduced detection rates of both semantic and non-semantic watermarking schemes while remaining within perceptual constrained bounds. 

\subsection{Research Motivation and Contributions}
UnMarker introduced two separate attacks: a non-semantic attack for removing high-frequency watermarks and a semantic attack for removing low-frequency watermarks. However, its evaluation considered images containing one watermark at a time.. This motivates the investigation of whether complementary watermarks types can be embedded simulatenously to counter the UnMarker attacks. Considering this, removing only one watermark is insufficient.

The objective of this study is to investigate Sequential Semantic
Non-Semantic Watermarking (SSNSW) by combining a non-semantic and a
semantic watermark within the same image. The study addresses the
following research questions: (1) Does the watermark embedding order influence the robustness of the SSNSW scheme? (2) Does the second embedded watermark affect the decodability of the first? (3) Does the order of the semantic and non-semantic UnMarker attacks affect watermark-removal performance?

To answer these questions, the experiments are organized into three
phases. Phase~1 establishes the baseline performance of HiDDeN and
StegaStamp under the corresponding non-semantic and semantic UnMarker
attacks. Phase~2 evaluates dual watermarking in both embedding orders:
HiDDeN followed by StegaStamp and StegaStamp followed by HiDDeN. Phase~3
applies both UnMarker attacks sequentially to the dual-watermarked images,
resulting in four experiments: (1) HiDDeN $\rightarrow$ StegaStamp
$\rightarrow$ semantic attack $\rightarrow$ non-semantic attack;
(2) HiDDeN $\rightarrow$ StegaStamp $\rightarrow$ non-semantic attack
$\rightarrow$ semantic attack; (3) StegaStamp $\rightarrow$ HiDDeN
$\rightarrow$ semantic attack $\rightarrow$ non-semantic attack; and
(4) StegaStamp $\rightarrow$ HiDDeN $\rightarrow$ non-semantic attack
$\rightarrow$ semantic attack.

HiDDeN and StegaStamp are used as the non-semantic and semantic watermarking schemes, respectively. After each embedding and attack, the bit accuracy of both watermarks are measured  to emsire they remain detectable. LPIPS is also recorded to evaluate perceptual cost of each watermarking and removal scheme. This
three-phase design enables the study of baseline attack effectiveness,
cross-watermark interference, embedding-order robustness, attack-order
effects, and whether sequential attacks are required to disrupt both
watermarks.

To the best of our knowloedge, bo study has systematically explored the combination of high- and low-frequency watermarks within a single image with sequential attacks designed to remove them. Existing works  mostly evaluate watermarking schemes and removals for single-layer watermarks. Therefore, this study contributes by extending watermark removal beyond isolated watermarking and removal schemes toward multi-layered watermarking and attack piplelines.

%% file: sections/03_methodology.tex
\section{Methodology}\label{chap:methodology}
This section describes the watermarking and watermark-removal
methods, datasets and preprocessing, evaluation metrics, and the
three-phase experimental design. It also presents the implementation settings and modifications used to extend the original methods to sequential watermarking and watermark removal methods.

\subsection{Selected Watermarking and Removal Methods}
Recent watermarking and watermark-removal methods can be computationally intensive, making an exhaustive evaluation of all available methods outside the scope of this study. 
Hence, HiDDeN and StegaStamp, one state-of-the-art non-semantic and semantic watermarking schemes, respectively, are selected for the experiments in this study. 

\paragraph{HiDDeN.}
HiDDeN \cite{zhu2018hidden} pioneered the use of end-to-end
deep-learning-based watermarking and serves as a building block for
several later watermarking schemes. HiDDeN uses a neural network to embed its signature (binary message) in images such that the image remains imperceptible. Yet, the HiDDeN decoder can recover the signal even after it undergoes common distortions. 

\begin{figure}[!t]
    \centering
    \includegraphics[width=1.0\linewidth]{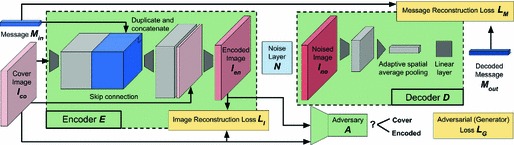}
    \caption{Overview of the HiDDeN watermarking framework. Source: adapted from \cite{zhu2018hidden}.}
    \label{fig:hidden_overview}
\end{figure}

The overview of HiDDeN is shown in Figure \ref{fig:hidden_overview}. The architecture can be broken down to four main components: The \textbf{Encoder}, a \textbf{Noise Layer}, the \textbf{Decoder}, and an \textbf{Adversarial Discriminator}. The Encoder, $E_{\theta}$ is tasked with encoding the message, $M_{\mathrm{in}}$, within the original image $I_{\mathrm{co}}$ to produce $I_{\mathrm{en}}$. The Noise Layer $N$ adds random noise mimicking real-life transformations of the watermarked image to produce $I_{\mathrm{no}}$. The Decoder $D_{\phi}$ retrieves the message from the distorted watermarked image $I_{\mathrm{no}}$. The Encoder and Decoder are jointly optimized through $\mathcal{L}_{E,D}$, which combines message-reconstruction loss, image-distortion loss, and adversarial loss. Finally, the Adversarial Discriminator $A_{\gamma}$ distinguishes between original and encoded image $\mathcal{L}_{A}$, encouraging encoder to produce watermarked images that remain similar to original image. The complete HiDDeN pipeline and its training objectives are expressed
compactly in \ref{eq:hidden_compact}.

\begin{equation}
\begin{aligned}
I_{\mathrm{en}} &= E_{\theta}(I_{\mathrm{co}},M_{\mathrm{in}}),\quad
I_{\mathrm{no}}=N(I_{\mathrm{co}},I_{\mathrm{en}}),\quad
M_{\mathrm{out}}=D_{\phi}(I_{\mathrm{no}}),\\
\mathcal{L}_{E,D}
&=\frac{\lambda_M}{L}\|M_{\mathrm{in}}-M_{\mathrm{out}}\|_2^2
+\lambda_I\|I_{\mathrm{co}}-I_{\mathrm{en}}\|_2^2
+\lambda_G\log\!\left(1-A_{\gamma}(I_{\mathrm{en}})\right),\\
\mathcal{L}_{A}
&=-\left[\log\!\left(1-A_{\gamma}(I_{\mathrm{co}})\right)
+\log\!\left(A_{\gamma}(I_{\mathrm{en}})\right)\right].
\end{aligned}
\label{eq:hidden_compact}
\end{equation}

Following the operational taxonomy adopted by UnMarker, HiDDeN is non-semantic and hides its signature in the high-frequency regions of the image. Hence, HiDDeN is selected as the non-semantic watermarking scheme for this study.

\paragraph{StegaStamp.}
\begin{figure}[!t]
    \centering
    \includegraphics[width=1.0\linewidth]{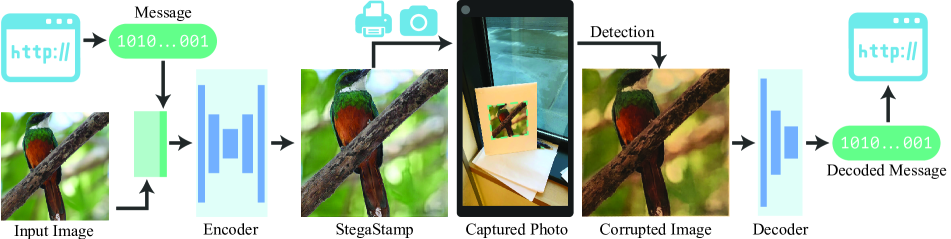}
    \caption{Working mechanism of StegaStamp, adapted from \cite{tancik2020stegastamp}.}
    \label{fig:stega_overview}
\end{figure}
StegaStamp \cite{tancik2020stegastamp} is built upon the architecture of HiDDeN with noticeable changes that make the watermarking feasible for printed images. The new functions extend the noise layer with differentiable transformations that are associated with printing, image capture and geometric distortions. Following the taxonomy adopted by UnMarker, StegaStamp is treated as a semantic watermarking scheme associated with lower-frequency watermark characteristics. StegaStamp is the only watermarking scheme that supports physical printed watermarking. Hence, due to its architectural similarity with HiDDeN and support for physical print--camera transformations, StegaStamp is selected as the semantic watermarking scheme for this study. 

The working mechanism of StegaStamp is shown in
Fig.~\ref{fig:stega_overview}. As summarized in
Eq.~\ref{eq:stegastamp_compact}, noise simulation $\mathcal{T}$ is applied between the Encoder $E_{\theta}$ and
the decoder $D_{\phi}$ so that the embedded message can be recovered after simulated printing camera distortions. StegaStamp additionally uses LPIPS and geometric reconstruction losses to preserve perceptual quality and improve robustness to geometric transformations.

\begin{equation}
M_{\mathrm{out}}
=
D_{\phi}\!\left(
\mathcal{T}\!\left(
E_{\theta}(\mathbf{x},M_{\mathrm{in}})
\right)\right),
\label{eq:stegastamp_compact}
\end{equation}

\paragraph{UnMarker.}
The UnMarker\cite{kassis2025unmarker} is then introduced for removing these watermarks. UnMarker is a black-box optimization algorithm that can remove semantic and non-semantic watermarks without any prior knowledge of the watermarking scheme. This makes UnMarker unique and the authors considered it the first practical attack on semantic watermarks. The goal of UnMarker is to produce an output image $x_{\text{unmark}}$ that is perceptually close to the input watermarked image $x'$ such that the decoder fails to read the hidden message $m$. As discussed from LPIPS basics, here the visual difference $l_d$ refers to visual difference between the watermarked image $x'$ and the output image $x_{\text{unmark}}$ and they must be less than a certain threshold $t_{ld}$. Therefore, the objective is shown in Equation \ref{eq:unmarker_obj}. 

\begin{equation}
\begin{aligned}
x' &\rightarrow x_{\text{unmark}},\\
l_d\!\left(x',x_{\text{unmark}}\right) &\le t_{ld},\\
F_c\!\left(x_{\text{unmark}}\right) &\neq
F_c\!\left(x'\right).
\end{aligned}
\label{eq:unmarker_obj}
\end{equation}

For successful implementation, the UnMarker assumes the methods we deduced earlier in HiDDeN and StegaStamp. That is, for a watermark to be robust against distortions, they must be embedded in specific spectral amplitudes (frequency bands). They call these the universal spectral carrier. Therefore, these assumptions led to two types of watermarking schemes. The non-semantic high-frequency attack and the semantic low-frequency attacks.

\textit{Non-semantic attack.}
Non-semantic watermarking schemes primarily inject watermarks in the
global spectral amplitudes, which are associated with high frequency characteristics. Direct Fourier Loss (DFL) measures the band-wise spectral differences between the watermarked image $x_w$ and attacked image $x_{\mathrm{um}}$, which UnMarker maximizes to disrupt the watermark. To ensure visual degradation is minimal, LPIPS $\ell_p(\cdot,\cdot)$ constrains perceptual deviation using the threshold $t_p$, while the $\ell_2$ norm threshold $t_2$ prevents the perturbation from being concentrated in only a small number of high-amplitude pixels. The attacked image $x_{\mathrm{um}}$ is obtained by optimizing DFL while penalizing any violation of the LPIPS and $\ell_2$ thresholds as expressed in Eq.\ref{eq:unmarker_nonsemantic}. Watermarks like HiDDeN in the high-frequency bands are significantly changed, reducing the decoders ability to recover the encoded message.

\begin{equation}
\begin{aligned}
x_{\mathrm{um}} &= x_w+\delta^\star,\\
\delta^\star
&=\arg\min_{\delta}\Big[
c_p\,\mathrm{ReLU}\!\left(
\ell_p(x_w+\delta,x_w)-t_p\right)\\
&\quad+c_2\,\mathrm{ReLU}\!\left(
\|x_w+\delta-x_w\|_2-t_2\right)
-\mathrm{DFL}(x_w+\delta,x_w)
\Big].
\end{aligned}
\label{eq:unmarker_nonsemantic}
\end{equation}

\textit{Semantic attack.}
Semantic watermarking schemes like StegaStamp embed watermarks into the low-frequency components of the image. As low-frequency watermarks are woven into the image textures with smooth transitions, the pixel-by-pixel manipulation of a non-semantic attack may fail to remove the watermark or introduce large visual artifacts. Therefore, UnMarker proposes a semantic attack in which groups of neighbouring pixels are manipulated simultaneously using optimizable filters to introduce spatially smooth perturbations. As shown in Eq.~\ref{eq:unmarker_semantic}, the filtered image is
obtained by applying $T$ optimizable filters with learnable parameters
$\theta$ to the watermarked image $x_w$ with additive perturbation
$\delta$. Each filtered pixel is calculated as a combination of its neighbours, producing spatially smooth intensity changes across the image. This disrupts low-frequency regions and reduce the decoder's ability to read the watermark.

DFL introduced in non-semantic attacks maximizes pixel-wise spectral changes, but it does not apply to neighbourhood-level perturbations. 
Hence, UnMarker introduces Filter Regularization Loss (FRL) and Mean Pool Loss (MPL). FRL  compares each pixel to its neighbours and penalizes large deviations as these outliers represent visual artifacts. However, FRL is limited to local medians; when multiple pixels are perturbed together, the local median may also change, which allows the watermark to remain intact. MPL addresses this limitation by comparing patch averages within a $L\times P$ ($5\times 5$) window. UnMarker maximizes $\mathrm{MPL}_{L\times P}(x_w,x)$ \ref{eq:unmarker_semantic}, forcing the output patches to deviate from those of the watermarked image. Together, MPL disrupts low-frequency patterns while FRL prevents visual artifacts and uniform brightness shifts. The objective in Eq.~\ref{eq:unmarker_semantic} jointly optimizes $\theta$ and $\delta$, penalizes violations of the LPIPS perceptual threshold and large $\ell_2$ deviations, and maximizes the MPL and FRL terms.

\begin{equation}
\begin{aligned}
x_{\theta,\delta}
&=\widehat{\mathrm{OF}}_{\theta}^{(T)}(x_w+\delta),\\
(\theta^\star,\delta^\star)
&=\arg\min_{\theta,\delta}\Big[
c_{\ell_{\mathrm{DVL}}}
\operatorname{ReLU}\!\left(
\ell_{\mathrm{DVL}}(x_w,x_{\theta,\delta})
-t_{\ell_{\mathrm{DVL}}}\right)\\
&\quad+\lambda_{2}
\left\|x_{\theta,\delta}-x_w\right\|_2
-\mathrm{MPL}_{5\times5}(x_w,x_{\theta,\delta})\\
&\quad-\lambda_{\mathrm{FRL}}
\mathrm{FRL}(x_{\theta,\delta})
\Big],\\
x_{\mathrm{um}}^{s}
&=\widehat{\mathrm{OF}}_{\theta^\star}^{(T)}
(x_w+\delta^\star).
\end{aligned}
\label{eq:unmarker_semantic}
\end{equation}

\subsection{Datasets used}
To evaluate the watermarking schemes and UnMarker effectively, datasets with natural scenes and structured facial image needs to be considered. Hence, for this study, two widely-used datasets, MS-COCO and CelebA-HQ, are considered. MS-COCO \cite{lin2014coco} is a high-quality dataset that contains complex real-world scenes with detailed backgrounds and diverse objects. The diversity enables the robustness of watermarking schemes to be evaluated across heterogeneous visual content. CelebA-HQ\cite{liu2015celeba} consist of high-quality facial imaging containing smooth gradients and facial patterns.These characteristics allow watermark behaviour to be evaluated in both smooth low-frequency regions, such as lighting, shading, and colour, and high-frequency regions, such as hair and eyelashes. Facial images also provide a structured image domain relevant to deepfake-related applications.

For this study, $50$ images are sampled. $25$ from MS-COCO and $25$ from CelebA-HQ. The computationally constrained evaluation scale follows the compact experiment scale of UnMarker \cite{kassis2025unmarker}, which evaluates each watermarking
system on 100 randomly selected images. The images are randomly sampled using a fixed seed to ensure reproducibility. They are then resized and cropped to a fixed resolution compatible with both watermarking schemes. The same set of images is used for all watermarking and attack and defense configurations to ensure consistency and a fair comparison.

\subsection{Evaluation criteria} \label{sec:evaluation-criteria}

WAVES Benchmark\cite {an2024waves} standardized metrics for future watermark attacks. The study finds that 'rinsing' and other techniques can remove stubborn semantic watermarks like TRW and StegaStamp for higher distortion within the image. Furthermore, WAVES mention that the success metrics of watermarking must also ensure image quality. This benchmarking is key to future watermark attack schemes.

In this study, efficiency of watermark removal is evaluated using two criteria: (1) \emph{decoding performance} via bit accuracy, and (2) \emph{visual fidelity} via perceptual similarity. The two methods are discussed in further detail below.

\subsubsection{Bit accuracy} \label{subsec:bit-accuracy}

Defensive watermarking methods consist of embedding a binary message
$m$ into an image $x$ to produce the watermarked image $x'$. The
watermark perturbation $\Delta x$ can be decoded by the decoder
$D(\cdot)$ to recover the binary message $m$. During the experiments,
the decoder outputs bit probabilities $p_j(I)$ and uses a threshold of
$0.5$ to make binary decisions. Bit Accuracy (BA) measures the success
rate of the decoder and is defined as

\begin{equation}
\hat{m}_j(I)=\mathbb{I}\!\left[p_j(I)\geq0.5\right],
\qquad
\mathrm{BA}(I)=\frac{1}{L}\sum_{j=1}^{L}
\mathbb{I}\!\left[\hat{m}_j(I)=m_j\right],
\label{eq:bit-accuracy}
\end{equation}

where $L$ is the message length and $\mathbb{I}[\cdot]$ is the
indicator function. The formula returns the proportion of the message
that the decoder can retrieve in the range $[0,1]$, where $1$
corresponds to $100\%$ of the message being recovered.

In the experiments, Bit Accuracy is evaluated in three stages:
$\mathrm{BA}_{\mathrm{orig}}=\mathrm{BA}(x)$,
$\mathrm{BA}_{\mathrm{wm}}=\mathrm{BA}(x')$, and
$\mathrm{BA}_{\mathrm{att}}=\mathrm{BA}(x_{\mathrm{att}})$.
Firstly, $\mathrm{BA}_{\mathrm{orig}}$ acts as the baseline sanity
check. Due to decoder bias, this value is crucial for evaluating the
failure of the decoder's performance. Secondly,
$\mathrm{BA}_{\mathrm{wm}}$ shows whether the watermark is functioning
appropriately and should ideally be close to $1$. Finally,
$\mathrm{BA}_{\mathrm{att}}$ quantifies the effectiveness of the attack
in removing the watermark and should ideally be close or equal to
$\mathrm{BA}_{\mathrm{orig}}$.

\subsubsection{LPIPS Similarity}
\label{subsec:lpips}

The constraint of watermark removal is to ensure that the attacked image remains visually as close as possible to the watermarked image while message recovery is disrupted. Hence, image fidelity is evaluated using Learned Perceptual Image Patch Similarity (LPIPS). LPIPS measures the perceptual distance between two images using CNN-based feature representations. For the watermarked image $x'$ and attacked output
$x_{\mathrm{att}}$, LPIPS is defined in Eq. \ref{eq:lpips-main}.

\begin{equation}
\begin{aligned}
\mathrm{LPIPS}(x',x_{\mathrm{att}})
&=
\sum_{\ell}
\frac{1}{H_{\ell}W_{\ell}}
\sum_{h,w}
\left\|
w_{\ell}\odot
\Delta\hat{f}_{\ell,hw}
\right\|_2^2,\\
\Delta\hat{f}_{\ell,hw}
&=
\hat{f}_{\ell}(x')_{hw}
-
\hat{f}_{\ell}(x_{\mathrm{att}})_{hw}.
\end{aligned}
\label{eq:lpips-main}
\end{equation}

where $\hat{f}_{\ell}(\cdot)$ denotes the channel-normalized feature
activation at layer $\ell$, $w_{\ell}$ contains the learned channel-wise weights, and $H_{\ell}$ and $W_{\ell}$ are the spatial dimensions of the corresponding feature map. Smaller LPIPS values indicate greater
perceptual similarity. UnMarker uses LPIPS as a perceptual constraint during optimization, while this study reports it as the image-fidelity metric. An attack is considered effective when $\mathrm{BA}_{\mathrm{att}}$ approaches the clean-image
baseline $\mathrm{BA}_{\mathrm{orig}}$ while
$\mathrm{LPIPS}(x',x_{\mathrm{att}})$ remains low.

\subsection{Experimental design}
The objective of this study is to evaluate the effects of combining two
watermarks, the robustness of the watermarks under UnMarker attacks, and the impact of embedding and attack order. The same benchmark of $N=50$ images from MS-COCO and CelebA-HQ is used. The two datasets contain objects and faces to ensure that the results do not become domain-specific. Results are reported together and separately for each dataset. The experiments are divided into three phases.

\paragraph{Phase 1: Individual watermark and attack baselines.}
The effectiveness of the semantic and non-semantic UnMarker attacks is
first evaluated. HiDDeN is tested against the non-semantic attack, while StegaStamp is tested against the semantic attack. The phase provides the baseline results of the study.

\begin{figure}[!t]
    \centering
    \includegraphics[width=1.0\linewidth]{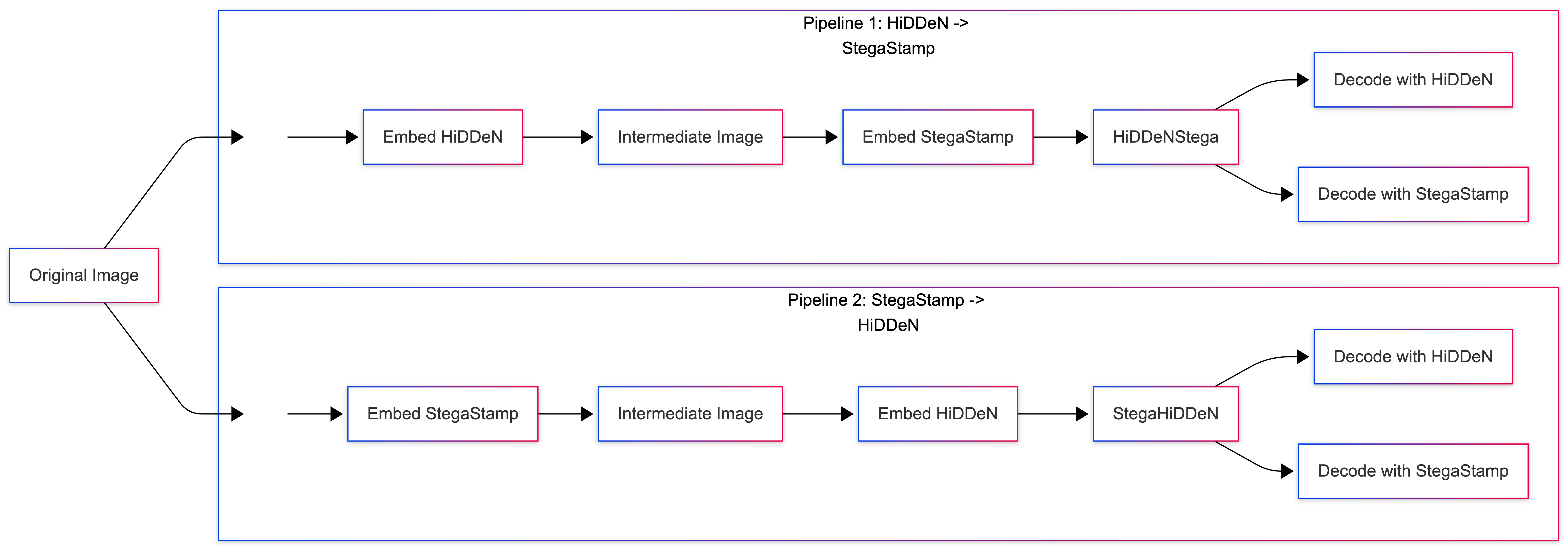}
    \caption{Dual-watermark embedding pipelines used in Phase 2.}
    \label{fig:phase2_pipeline}
\end{figure}

\paragraph{Phase 2: Dual-watermark embedding and interference study.}
For Phase 2, dual watermarking schemes are implemented to provide images
with two layers of defensive watermarks. This experiment studies the
impact of embedding sequence and evaluates the interference caused when
the second watermark overlaps the first. A bird's-eye view of this phase
is shown in Fig.~\ref{fig:phase2_pipeline}.

Two dual-watermarking pipelines are applied to the benchmark dataset of
$N=50$ images $\{x_i\}_{i=1}^{N}$. In HiDDeNStega, each image $x_i$ is
watermarked with HiDDeN to produce $x_i^{H}$ and then with StegaStamp to
produce $x_i^{HS}$. The resulting dataset $X^{HS}$ is referred to as the
\textbf{HiDDeNStega dataset}. In StegaHiDDeN, each image is watermarked
with StegaStamp to produce $x_i^{S}$ and then with HiDDeN to produce
$x_i^{SH}$. The resulting dataset $X^{SH}$ is referred to as the
\textbf{StegaHiDDeN dataset}.

After each embedding stage, the Bit Accuracy of HiDDeN and StegaStamp is evaluated alongside LPIPS. These measurements evaluate how much the
second embedding affects the first watermark and whether the images
remain perceptually similar after dual watermarking.

\begin{figure}[!t]
    \centering
    \includegraphics[width=1.0\linewidth]{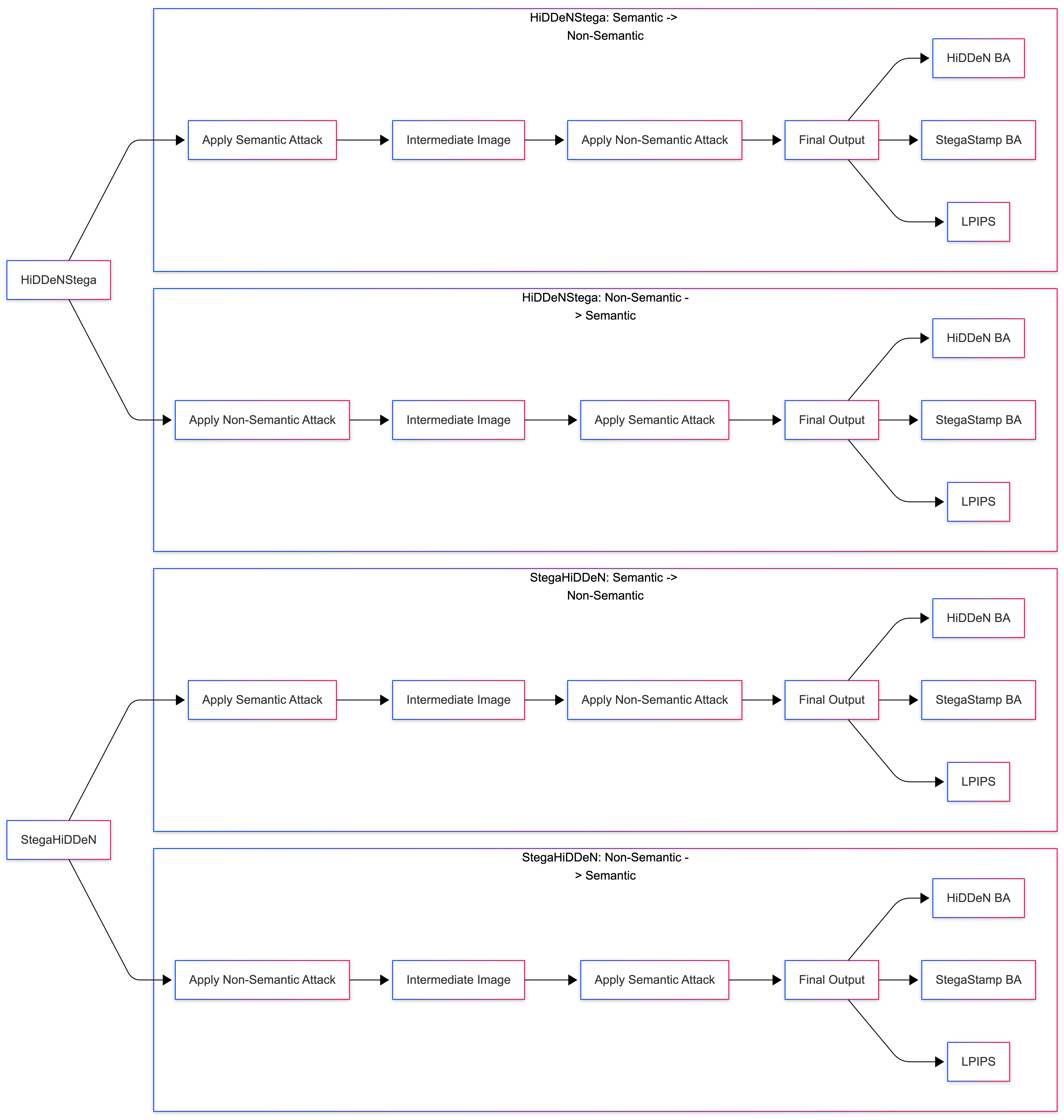}
    \caption{Sequential attack pipelines used in Phase 3.}
    \label{fig:phase3_pipeline}
\end{figure}

\paragraph{Phase 3: Dual-watermark removal and attack-order study.}
This phase evaluates the robustness of the dual-watermarking schemes and the effectiveness of UnMarker when both types of watermarks are present. The main outcome is to determine whether both attacks are necessary and, if so, which attack sequence is more effective. A bird's-eye view of
Phase 3 is shown in Fig.~\ref{fig:phase3_pipeline}.

For each dual-watermarked dataset, $X^{HS}$ and $X^{SH}$, two sequential attack orders are applied: non-semantic followed by semantic, and semantic followed by non-semantic. This produces four attack pipelines, covering both attack orders for both dual-watermarked datasets.

For each embedding and attack stage, the Bit Accuracy of the HiDDeN and
StegaStamp decoders and LPIPS are recorded. This enables analysis of the drop in decoder performance after each attack, the effectiveness of each attack sequence, whether sequential attacks are necessary, and their impact on perceptual similarity.

%% file: sections/04_results.tex
\section{Results}\label{chap:results}

The results are broken down into three phases based on the experimental
design of the study. First, the results of the baseline watermarking
schemes are provided against the basic UnMarker attacks. Then, Phase 2
shows the results of the dual watermarking schemes. Finally, Phase 2
results are used in Phase 3 to evaluate the SSNSW schemes against the
sequential UnMarker attacks.

\subsection{Phase 1: Baseline study}
Phase 1 is further broken down into two experiments: HiDDeN watermarking scheme being tested by the non-semantic attack, followed by the semantic attack on StegaStamp.

\begin{figure}[!t]
    \centering
    \includegraphics[width=\linewidth]{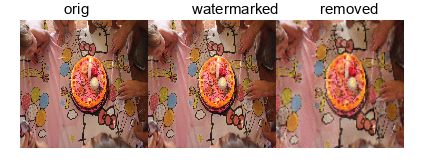}
    \caption{Example with \textbf{highest} attacked bit accuracy
    ($\mathrm{BA}_{\mathrm{att}}$) under the non-semantic attack on the
    HiDDeN-embedded dataset.}
    \label{fig:base_hidden_high}
\end{figure}

\begin{figure}[!t]
    \centering
    \includegraphics[width=\linewidth]{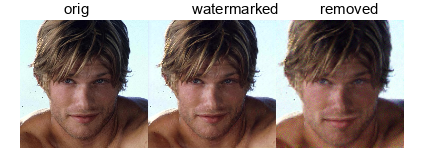}
    \caption{Example with \textbf{lowest} attacked bit accuracy
    ($\mathrm{BA}_{\mathrm{att}}$) under the non-semantic attack on the
    HiDDeN-embedded dataset.}
    \label{fig:base_hidden_low}
\end{figure}

Table \ref{tab:hidden_base_nonsemantic_summary} shows the results on the benchmark dataset of the HiDDeN watermarking scheme followed by the UnMarker attack. On the original dataset, the HiDDeN decoder bit accuracy of HiDDeN was quite low at approximately $0.58$ for both datasets. Then the watermarking worked perfectly and all the images are marked at $1.00$, which is fully decodable. The non-semantic UnMarker attacks reduced the decoder readability close to original baseline, while LPIPS remained low.  However, some images retained traces of HiDDeN with a decoder accuracy of $0.70$. The attack produced a lower average bit accuracy on COCO, while CelebA had a slightly higher average LPIPS score. Figures~\ref{fig:base_hidden_high} and \ref{fig:base_hidden_low} show the images with the highest and lowest attacked bit accuracies respectively.

\begin{table}[!t]
\centering
\small
\caption{Results for base HiDDeN under the Non-Semantic Attack.}
\label{tab:hidden_base_nonsemantic_summary}
\begin{tabular}{lcccc}
\toprule
Subset & $\mathrm{BA}_{\text{orig}}$ & $\mathrm{BA}_{\text{wm}}$ & $\mathrm{BA}_{\text{att}}$ & LPIPS \\
\midrule
COCO (1--25)   & 0.581 & 1.000 & 0.549 & 0.070 \\
CelebA (26--50)& 0.584 & 1.000 & 0.567 & 0.082 \\
Overall (1--50)& 0.583 & 1.000 & 0.558 & 0.076 \\
\bottomrule
\end{tabular}
\end{table}

\begin{figure}[!t]
     \centering
     \includegraphics[width=\linewidth]{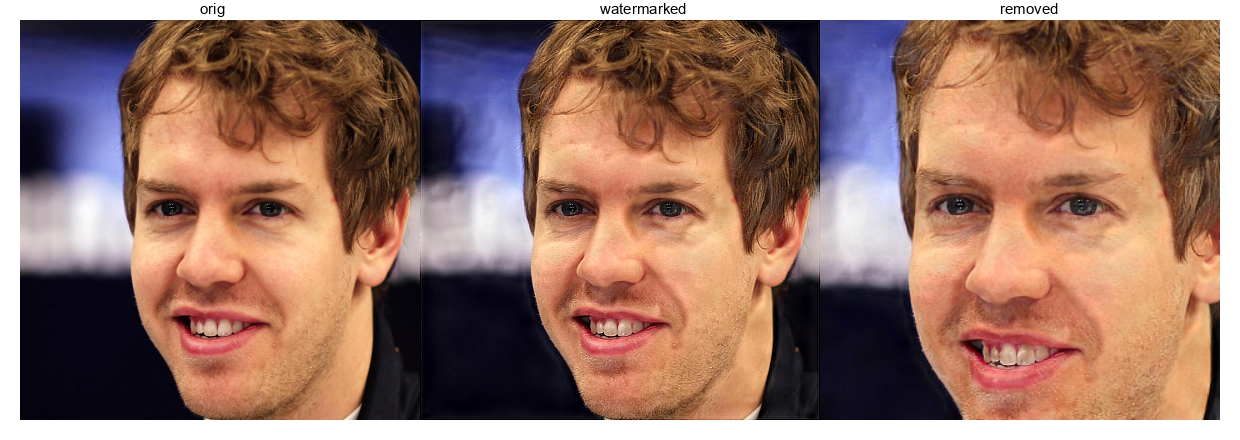}
     \caption{Example with \textbf{lowest} attacked bit accuracy ($\mathrm{BA}_{\text{att}}$) under the semantic attack on the base StegaStamp dataset.}
     \label{fig:base_stega_low}
\end{figure}

\begin{figure}[!t]
     \centering
     \includegraphics[width=\linewidth]{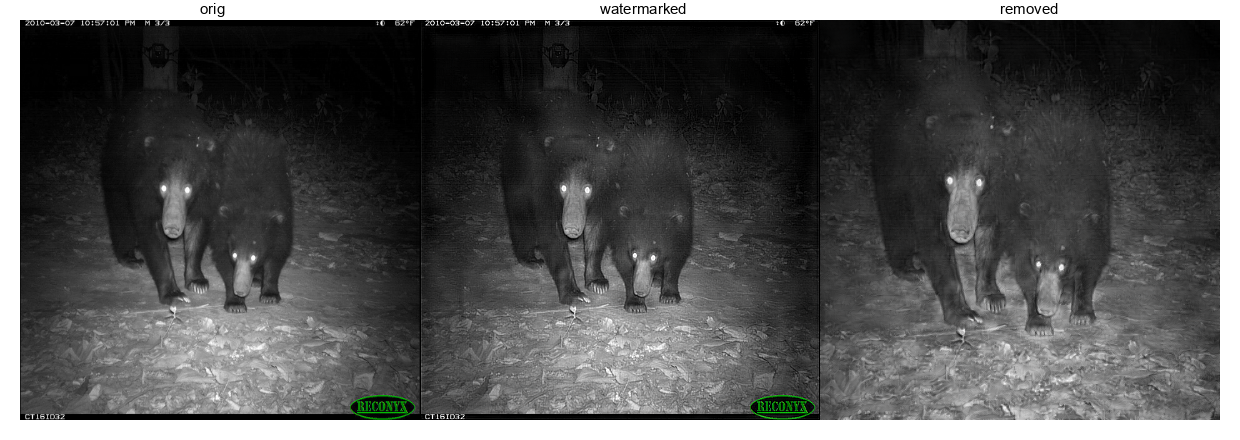}
     \caption{Example with \textbf{highest} attacked bit accuracy ($\mathrm{BA}_{\text{att}}$) under the semantic attack on the base StegaStamp dataset.}
     \label{fig:base_stega_high}
\end{figure}

Table~\ref{tab:stega_base_semantic_summary} displays the results of the
semantic attack on the StegaStamp-embedded dataset. Similar to
HiDDeN, StegaStamp ensured full decodability with perfect bit accuracy.
The semantic attack reduced the effectiveness of StegaStamp, although
some traces of the watermark remained. An average bit accuracy below
$0.6$ suggests that the majority of the watermark was removed. The
attack produced a lower average bit accuracy on the CelebA dataset.
Some images retained a high bit accuracy of $0.690$. The differences are shown in Fig.~\ref{fig:base_stega_low}, with the lowest attacked bit accuracy, and Fig.~\ref{fig:base_stega_high}, with the highest attacked bit accuracy.

\begin{table}[!t]
\centering
\small
\caption{Results for base StegaStamp under the Semantic Attack.}
\label{tab:stega_base_semantic_summary}
\begin{tabular}{lcccc}
\toprule
Subset & $\mathrm{BA}_{\text{orig}}$ & $\mathrm{BA}_{\text{wm}}$ & $\mathrm{BA}_{\text{att}}$ & LPIPS \\
\midrule
COCO (1--25)    & 0.496 & 0.999 & 0.602 & 0.067 \\
CelebA (26--50) & 0.520 & 1.000 & 0.563 & 0.072 \\
Overall (1--50) & 0.508 & 1.000 & 0.583 & 0.070 \\
\bottomrule
\end{tabular}
\end{table}

\subsection{Phase 2: Watermarking Sequence Study}
\label{subsec:phase2-results}
Phase two of the experiments test the proposed dual watermarking study.
HiDDeN and StegaStamp are applied sequentially to the benchmark dataset in both sequences. HiDDeNStega represents HiDDeN embedding followed by StegaStamp, while StegaHiDDeN represents StegaStamp embedding followed by HiDDeN.

\subsubsection{Phase 2.1: HiDDeNStega}
\begin{figure}[!t]
    \centering
    
    \subfloat[Original\label{fig:p2_hs_step0}]{%
\begin{minipage}[t]{0.3\linewidth}
        \centering
        \includegraphics[height=2.5cm,keepaspectratio]{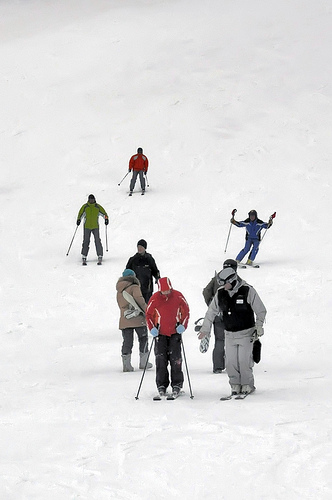}
\end{minipage}}
    \hfill
    \subfloat[HiDDeN watermarked\label{fig:p2_hs_step1}]{%
\begin{minipage}[t]{0.3\linewidth}
        \centering
        \includegraphics[height=2.5cm,keepaspectratio]{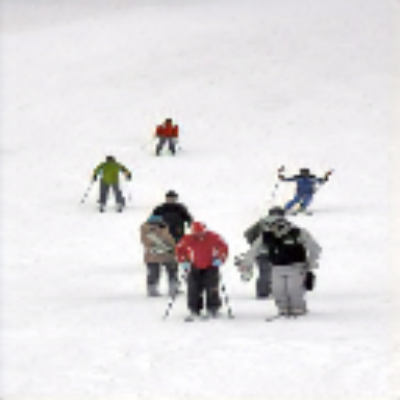}
\end{minipage}}
    \hfill
    \subfloat[HiDDeNStega watermarked\label{fig:p2_hs_step2}]{%
\begin{minipage}[t]{0.3\linewidth}
        \centering
        \includegraphics[height=2.5cm,keepaspectratio]{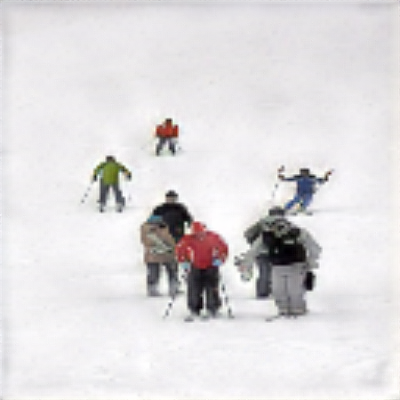}
\end{minipage}}
    
    \caption{HiDDeNStega watermarking pipeline (Phase 2): original image, after HiDDeN embedding, and after subsequent StegaStamp embedding.}
    \label{fig:p2_hiddensstega_pipeline}
\end{figure}

In this section, the dataset is first embedded with HiDDeN. Afterwards, StegaStamp is applied to the dataset to create the HiDDeNStega dataset. Sample result of the dual watermarked image is shown in Figure \ref{fig:p2_hiddensstega_pipeline}. Results from applying HiDDeN watermarking on the original dataset are provided in Table\ref{tab:hiddenstega_embedding_summary}. Results show that embedding HiDDeN alone is highly successful as it provides very low visual perception change while the HiDDeN watermark remains highly decodable. Results from applying StegaStamp on the HiDDeN embedded dataset to create the HiDDeNStega scheme is provided in Table \ref{tab:hiddenstega_embedding_summary}. This dataset is referred to as the HiDDeNStega dataset (HSD) going forward. 

The results show that, on average, the StegaStamp embeddings provide a much higher perception change than HiDDeN. This is consistent with the base experiments, where more significant distortion was produced with semantic watermarks. However, the important finding is that the decoder scores of HiDDeN still remain relatively perfect with a bit accuracy of over 97\% on average. Therefore, the theory confirms that dual watermarking schemes are applicable if they are embedded in different frequency domains of the image. Figure \ref{fig:p2_hiddensstega_pipeline} shows the three stages of the image during the HiDDeNStega watermarking scheme. The blurring effects of StegaStamp are slightly noticeable, while the HiDDeN watermarked images seem relatively similar. Despite this, the images can still be used to prevent scammers from pirating the image.

\begin{table}[!t]
\centering
\scriptsize
\setlength{\tabcolsep}{3pt}
\renewcommand{\arraystretch}{1.03}
\caption{Phase 2 results for the HiDDeN$\rightarrow$StegaStamp
embedding order. H and S denote HiDDeN and StegaStamp, respectively.}
\label{tab:hiddenstega_embedding_summary}
\begin{tabular}{@{}llccc@{}}
\toprule
Stage & Subset
& $\mathrm{BA}_{\mathrm{H}}$
& $\mathrm{BA}_{\mathrm{S}}$
& LPIPS \\
\midrule

\multirow{2}{*}{After H}
& COCO   & 0.998 & 0.503 & 0.038 \\
& CelebA & 0.994 & 0.523 & 0.052 \\
\midrule

\multirow{2}{*}{After S}
& COCO   & 0.976 & 0.982 & 0.142 \\
& CelebA & 0.986 & 0.981 & 0.103 \\
\bottomrule
\end{tabular}
\end{table}

\subsubsection{Phase 2.2: StegaHiDDeN}
\begin{figure}[!t]
    \centering
    
    \subfloat[Original\label{fig:p2_sh_step0}]{%
\begin{minipage}[t]{0.3\linewidth}
        \centering
        \vspace{0pt}
        \includegraphics[width=\linewidth]{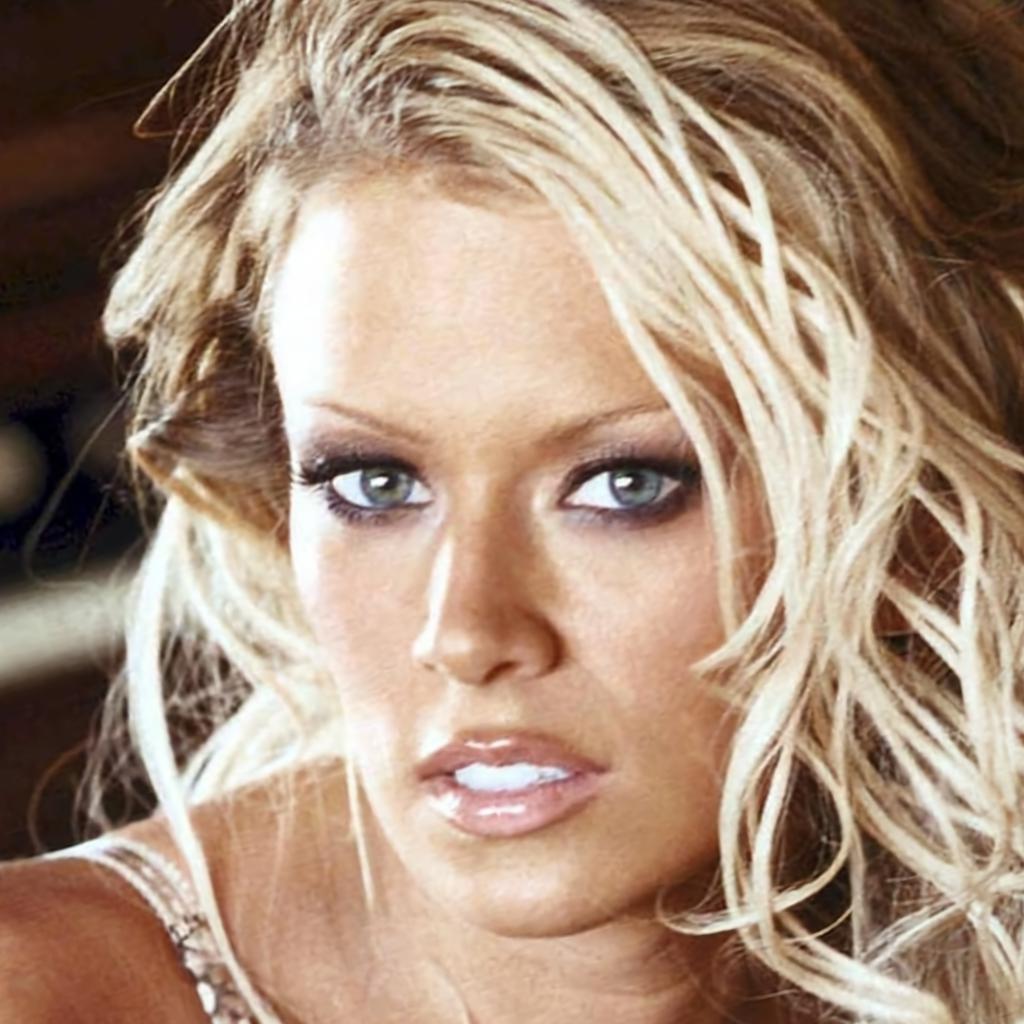}
\end{minipage}}
    \hfill
    \subfloat[StegaStamp watermarked\label{fig:p2_sh_step1}]{%
\begin{minipage}[t]{0.3\linewidth}
        \centering
        \vspace{0pt}
        \includegraphics[width=\linewidth]{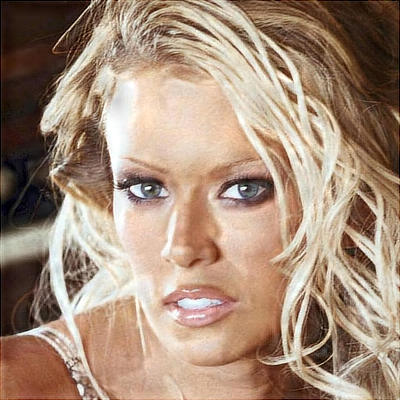}
\end{minipage}}
    \hfill
    \subfloat[StegaHiDDeN watermarked\label{fig:p2_sh_step2}]{%
\begin{minipage}[t]{0.3\linewidth}
        \centering
        \vspace{0pt}
        \includegraphics[width=\linewidth]{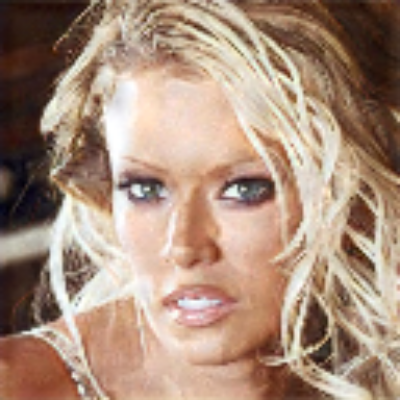}
\end{minipage}}
    
    \caption{StegaHiDDeN watermarking pipeline (Phase 2): original image, after StegaStamp embedding, and after subsequent HiDDeN embedding.}
    \label{fig:p2_stegahidden_pipeline}
\end{figure}

In this section, the original dataset is first embedded with StegaStamp. Afterwards, HiDDeN is applied on the StegaStamp embedded dataset to create the StegaHiDDeN dataset. Figure \ref{fig:p2_stegahidden_pipeline} shows a sample result of the dual watermarking process. Results from applying StegaStamp watermarking on the original dataset is provided in Table \ref{tab:stegahidden_embedding_summary}. As seen in the HiDDeNStega example, the perceptual difference is comparatively higher for StegaStamp embeddings. However, the decoder perfectly reads the watermark for StegaStamp. Results from applying HiDDeN on the Stega dataset to create the StegaHiDDeN scheme are provided in Table \ref{tab:stegahidden_embedding_summary}. This dataset is referred to as the StegaHiDDeN dataset (SHD) going forward. 

Once again, the theory is confirmed that sequential watermarking can be used as one watermark does not disrupt the other one. Both watermarks remain almost perfectly decodable regardless of the order they are used in.

\begin{table}[h]
\centering
\scriptsize
\setlength{\tabcolsep}{3pt}
\renewcommand{\arraystretch}{1.03}
\caption{Phase 2 results for the StegaStamp$\rightarrow$HiDDeN
embedding order. H and S denote HiDDeN and StegaStamp, respectively.}
\label{tab:stegahidden_embedding_summary}
\begin{tabular}{@{}llccc@{}}
\toprule
Stage & Subset
& $\mathrm{BA}_{\mathrm{H}}$
& $\mathrm{BA}_{\mathrm{S}}$
& LPIPS \\
\midrule

\multirow{2}{*}{After S}
& COCO   & 0.437 & 0.998 & 0.179 \\
& CelebA & 0.459 & 1.000 & 0.145 \\
\midrule

\multirow{2}{*}{After H}
& COCO   & 1.000 & 0.994 & 0.035 \\
& CelebA & 0.993 & 0.992 & 0.052 \\
\bottomrule
\end{tabular}
\end{table}

\subsection{Phase 3: Dual-watermark removal and attack-order
study}

\begin{figure}[h]
    \centering
    
    \subfloat[Dual watermarked\label{fig:p3_hs_base}]{%
\begin{minipage}[t]{0.3\linewidth}
        \centering
        \includegraphics[width=\linewidth]{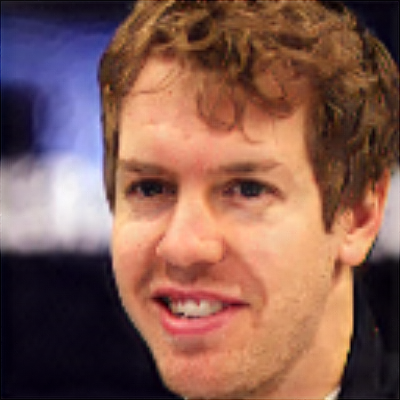}
\end{minipage}}
    \hfill
    \subfloat[After non-semantic\label{fig:p3_hs_ns}]{%
\begin{minipage}[t]{0.3\linewidth}
        \centering
        \includegraphics[width=\linewidth]{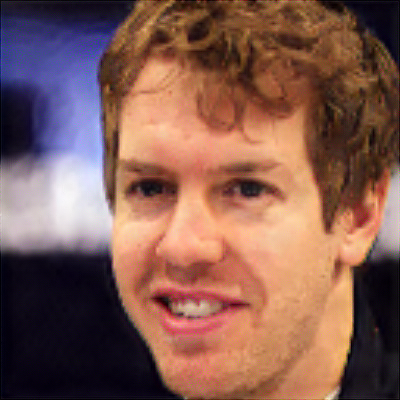}
\end{minipage}}
    \hfill
    \subfloat[After non-semantic $\rightarrow$ Semantic\label{fig:p3_hs_ns_s}]{%
\begin{minipage}[t]{0.3\linewidth}
        \centering
        \includegraphics[width=\linewidth]{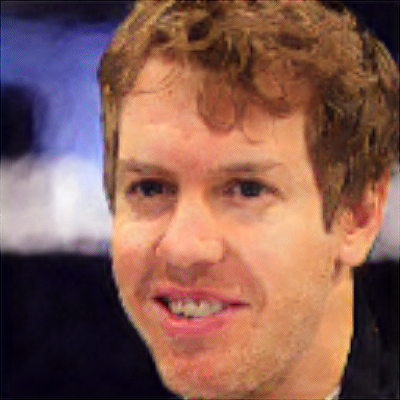}
\end{minipage}}
    
    \vspace{0.6em}
    
    \subfloat[Dual watermarked\label{fig:p3_hs_base_repeat}]{%
\begin{minipage}[t]{0.3\linewidth}
        \centering
        \includegraphics[width=\linewidth]{images/Experiments/P3_HS.png}
\end{minipage}}
    \hfill
    \subfloat[After semantic\label{fig:p3_hs_s}]{%
\begin{minipage}[t]{0.3\linewidth}
        \centering
        \includegraphics[width=\linewidth]{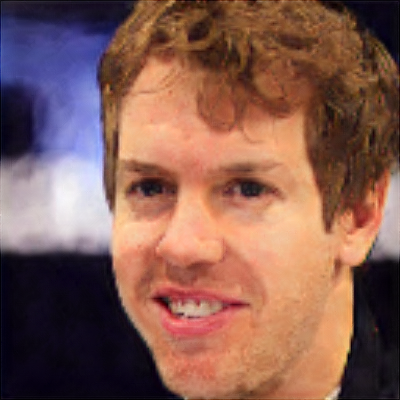}
\end{minipage}}
    \hfill
    \subfloat[After semantic $\rightarrow$ Non-semantic\label{fig:p3_hs_s_ns}]{%
\begin{minipage}[t]{0.3\linewidth}
        \centering
        \includegraphics[width=\linewidth]{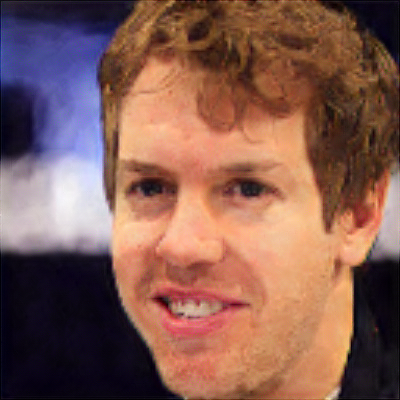}
\end{minipage}}
    
    \caption{HiDDeNStega attack pipelines (Phase 3). Top row: Non-semantic $\rightarrow$ Semantic. Bottom row: Semantic $\rightarrow$ Non-semantic.}
    \label{fig:p3_hs_attack_pipelines}
\end{figure}

\paragraph{HiDDeNStega.}
Tables \ref{tab:HS_NS_S_combined_summary} and \ref{tab:HS_S_NS_combined_summary} show the results of applying both
attack sequences to the HiDDeNStega dataset. The sample results are shown in Figure \ref{fig:p3_hs_attack_pipelines}. The non-semantic attack followed by the semantic attack reduces the bit accuracies of both watermarks. Similarly, applying the semantic attack first and the non-semantic attack afterwards also reduces StegaStamp and HiDDeN respectively, but traces of StegaStamp remain strong. 

\begin{figure}[h]
    \centering
    
    \subfloat[Dual watermarked\label{fig:p3_sh_base}]{%
\begin{minipage}[t]{0.3\linewidth}
        \centering
        \includegraphics[width=\linewidth]{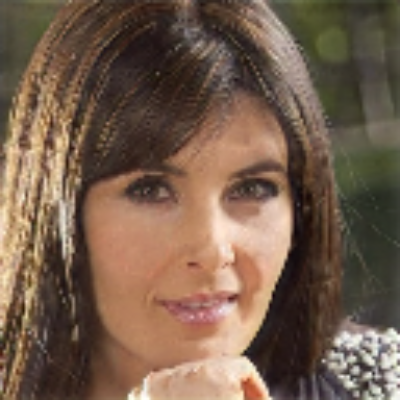}
\end{minipage}}
    \hfill
    \subfloat[After non-semantic\label{fig:p3_sh_ns}]{%
\begin{minipage}[t]{0.3\linewidth}
        \centering
        \includegraphics[width=\linewidth]{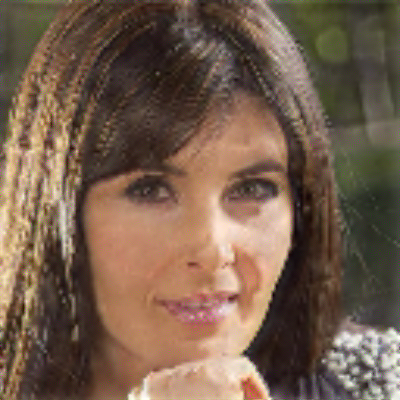}
\end{minipage}}
    \hfill
    \subfloat[After non-semantic $\rightarrow$ semantic\label{fig:p3_sh_ns_s}]{%
\begin{minipage}[t]{0.3\linewidth}
        \centering
        \includegraphics[width=\linewidth]{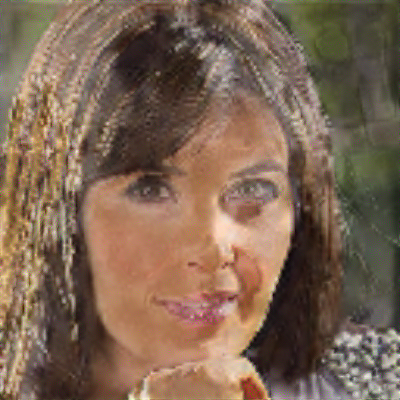}
\end{minipage}}
    
    \vspace{0.6em}
    
    \subfloat[Dual watermarked\label{fig:p3_sh_base_repeat}]{%
\begin{minipage}[t]{0.3\linewidth}
        \centering
        \includegraphics[width=\linewidth]{images/Experiments/P3_SH.png}
\end{minipage}}
    \hfill
    \subfloat[After semantic\label{fig:p3_sh_s}]{%
\begin{minipage}[t]{0.3\linewidth}
        \centering
        \includegraphics[width=\linewidth]{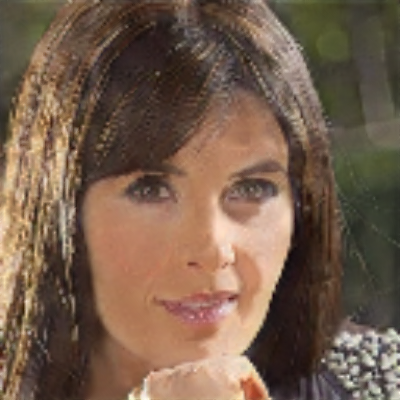}
\end{minipage}}
    \hfill
    \subfloat[After semantic $\rightarrow$ Non-semantic\label{fig:p3_sh_s_ns}]{%
\begin{minipage}[t]{0.3\linewidth}
        \centering
        \includegraphics[width=\linewidth]{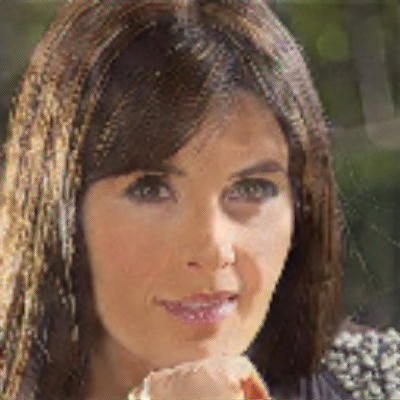}
\end{minipage}}
    
    \caption{StegaHiDDeN attack pipelines (Phase 3). Top row: Non-semantic $\rightarrow$ Semantic. Bottom row: Semantic $\rightarrow$ Non-semantic.}
    \label{fig:p3_sh_attack_pipelines}
\end{figure}

\paragraph{StegaHiDDeN.}
Results from applying both attack sequences to the StegaHiDDeN dataset
are shown in tables \ref{tab:SH_NS_S_combined_summary} and \ref{tab:SH_S_NS_combined_summary}. Sample results are shown in figure \ref{fig:p3_sh_attack_pipelines}. The results show that both sequence of attacks (NS followed by S and S followed by NS) can successfully reduce bit accuracies of the watermarks. However, it is to be noted that some traces of watermarks remain strongly, especially StegaStamp, the semantic watermark similar to the case observed above.

\begin{table}[!t]
\centering
\scriptsize
\setlength{\tabcolsep}{3pt}
\renewcommand{\arraystretch}{1.03}
\caption{Phase 3 results for \textit{HiDDeNStega} under the
Non-Semantic$\rightarrow$Semantic attack order. H and S denote
HiDDeN and StegaStamp, respectively.}
\label{tab:HS_NS_S_combined_summary}
\begin{tabular}{@{}llccc@{}}
\toprule
Stage & Subset
& $\mathrm{BA}_{\mathrm{H}}$
& $\mathrm{BA}_{\mathrm{S}}$
& LPIPS \\
\midrule

\multirow{2}{*}{After NS}
& COCO   & 0.584 & 0.954 & 0.009 \\
& CelebA & 0.614 & 0.960 & 0.010 \\
\midrule

\multirow{2}{*}{After S}
& COCO   & 0.554 & 0.720 & 0.081 \\
& CelebA & 0.593 & 0.657 & 0.085 \\
\bottomrule
\end{tabular}
\end{table}

\begin{table}[!t]
\centering
\scriptsize
\setlength{\tabcolsep}{3pt}
\renewcommand{\arraystretch}{1.03}
\caption{Phase 3 results for \textit{HiDDeNStega} under the
Semantic$\rightarrow$Non-Semantic attack order. H and S denote
HiDDeN and StegaStamp, respectively.}
\label{tab:HS_S_NS_combined_summary}
\begin{tabular}{@{}llccc@{}}
\toprule
Stage & Subset
& $\mathrm{BA}_{\mathrm{H}}$
& $\mathrm{BA}_{\mathrm{S}}$
& LPIPS \\
\midrule

\multirow{2}{*}{After S}
& COCO   & 0.858 & 0.671 & 0.087 \\
& CelebA & 0.909 & 0.613 & 0.080 \\
\midrule

\multirow{2}{*}{After NS}
& COCO   & 0.591 & 0.683 & 0.007 \\
& CelebA & 0.622 & 0.640 & 0.009 \\
\bottomrule
\end{tabular}
\end{table}

\begin{table}[!t]
\centering
\scriptsize
\setlength{\tabcolsep}{3pt}
\renewcommand{\arraystretch}{1.03}
\caption{Phase 3 results for \textit{StegaHiDDeN} under the
Non-Semantic$\rightarrow$Semantic attack order. H and St denote
HiDDeN and StegaStamp, respectively.}
\label{tab:SH_NS_S_combined_summary}
\begin{tabular}{@{}llccc@{}}
\toprule
Stage & Subset
& $\mathrm{BA}_{\mathrm{H}}$
& $\mathrm{BA}_{\mathrm{St}}$
& LPIPS \\
\midrule

\multirow{2}{*}{After NS}
& COCO   & 0.640 & 0.989 & 0.009 \\
& CelebA & 0.677 & 0.988 & 0.010 \\
\midrule

\multirow{2}{*}{After Sem}
& COCO   & 0.603 & 0.619 & 0.155 \\
& CelebA & 0.628 & 0.614 & 0.140 \\
\bottomrule
\end{tabular}
\end{table}

\begin{table}[!t]
\centering
\scriptsize
\setlength{\tabcolsep}{3pt}
\renewcommand{\arraystretch}{1.03}
\caption{Phase 3 results for \textit{StegaHiDDeN} under the
Semantic$\rightarrow$Non-Semantic attack order. H and St denote
HiDDeN and StegaStamp, respectively.}
\label{tab:SH_S_NS_combined_summary}
\begin{tabular}{@{}llccc@{}}
\toprule
Stage & Subset
& $\mathrm{BA}_{\mathrm{H}}$
& $\mathrm{BA}_{\mathrm{St}}$
& LPIPS \\
\midrule

\multirow{2}{*}{After Sem}
& COCO   & 0.880 & 0.581 & 0.144 \\
& CelebA & 0.914 & 0.580 & 0.128 \\
\midrule

\multirow{2}{*}{After NS}
& COCO   & 0.602 & 0.594 & 0.009 \\
& CelebA & 0.621 & 0.595 & 0.010 \\
\bottomrule
\end{tabular}
\end{table}

%% file: sections/05_discussion.tex
\section{Discussion}\label{chap:discussion}
This section discusses the performance of the individual attacks,
dual-watermark embedding, sequential watermark removal, and perceptual
quality. The limitations and future scope of the study are discussed
afterwards.

\subsection{Baseline Watermarking and Individual Attack Performance}

Phase 1 shows that both watermarking schemes are consistent and provide bit accuracies of $1.0$ before the attack. The corresponding  non-semantic and semantic attacks substantially reduce the bit accuracy of the watermarks; HiDDeN's bit accuracy drops from
$\mathrm{BA}_{\mathrm{wm}}=1.000$ to $\mathrm{BA}_{\mathrm{att}}=0.558$, is close to the original-image baseline of $\mathrm{BA}_{\mathrm{orig}}=0.583$.However, the per-image results show some variation, as certain images retain traces of the watermark with a bit accuracy of up to $0.70$ after the attack.

Similar results are seen when StegaStamp watermarked are subjected to the semantic attack. The bit accuracy drops from $1.000$ to $0.583$, approaching the original-image baseline of $0.508$. Even then, certain images contain traces of the watermark with a bit accuracy of up to $0.69$. but still not as strong as it was before the attacks. These findings can be studied further to determine why traces of watermarks remain.  These findings show that both attacks are
effective overall.

\begin{table}[!t]
\centering
\scriptsize
\setlength{\tabcolsep}{2.5pt}
\renewcommand{\arraystretch}{1.03}
\caption{Baseline results over COCO and CelebA. H and S denote
HiDDeN and StegaStamp; NS and Sem denote non-semantic and semantic
attacks, respectively.}
\label{tab:baseline_overall_summary}
\begin{tabular}{@{}llcccc@{}}
\toprule
WM & Attack
& $\mathrm{BA}_{\mathrm{orig}}$
& $\mathrm{BA}_{\mathrm{wm}}$
& $\mathrm{BA}_{\mathrm{att}}$
& LPIPS \\
\midrule
H  & NS  & 0.583 & 1.000 & 0.558 & 0.076 \\
S & Sem & 0.508 & 1.000 & 0.583 & 0.070 \\
\bottomrule
\end{tabular}
\end{table}

\subsection{Dual-Watermark Embedding and Embedding-Order Effects}

The second phase of experiments tests the proposed dual-watermarking
study. HiDDeN and StegaStamp are applied in both sequences.
HiDDeNStega represents HiDDeN embedding followed by StegaStamp,
while StegaHiDDeN represents the reverse order. Results from phase 2 in Tables \ref{tab:hiddenstega_embedding_summary} and \ref{tab:stegahidden_embedding_summary} support the hypothesis that two watermarks targeting different image characteristics can be embedded together without significant disruption to each other. For both sequences, the watermarks remain decodable, and the reduction in bit accuracy of the first watermark after embedding the second one is minimal.

HiDDeNStega achieves final bit accuracies of $0.981$ and $0.982$ for
HiDDeN and StegaStamp, respectively. Applying StegaStamp after HiDDeN
slightly decreases the HiDDeN bit accuracy, but both watermarks remain
almost fully decodable. The StegaHiDDeN results are more encouraging, with final bit accuracies of $0.997$ and $0.993$ for HiDDeN and StegaStamp, respectively. Here, HiDDeN's low-distortion properties result in only a small reduction in StegaStamp bit accuracy. Therefore, embedding with HiDDeN last creates slightly better decodability. Embedding order has a small effect on the final decoder performance. For the pair in this experiment, StegaStamp followed by HiDDeN provides slightly better final bit accuracies for both decoders.

\subsection{Sequential Watermark Removal and Attack-Order Effects}
Phase 3 evaluates the dual-watermarking schemes under multiple attacks.
The first question is whether they can withstand individual UnMarker
attacks and still carry traces of at least one watermark. Their
robustness is then tested further using sequential attacks on both the
HiDDeNStega and StegaHiDDeN schemes. 

\begin{table}[!t]
\centering
\scriptsize
\setlength{\tabcolsep}{2.2pt}
\renewcommand{\arraystretch}{1.02}
\caption{Phase 3 summary over COCO and CelebA. H$\rightarrow$S and
S$\rightarrow$H denote the HiDDeN--StegaStamp embedding orders;
NS and S denote non-semantic and semantic attacks, respectively.}
\label{tab:phase3_sequential_attacks_summary_compact}

\begin{tabular}{@{}lllccc@{}}
\toprule
Embed. & Attack order & Stage
& $\mathrm{BA}_{\mathrm{H}}$
& $\mathrm{BA}_{\mathrm{S}}$
& LPIPS \\
\midrule

\multirow{5}{*}{H$\rightarrow$S}
& --                         & WM       & 0.981 & 0.982 & 0.123 \\
& \multirow{2}{*}{NS$\rightarrow$S}
                             & After NS & 0.599 & 0.957 & 0.009 \\
&                            & After S  & 0.573 & 0.689 & 0.083 \\
& \multirow{2}{*}{S$\rightarrow$NS}
                             & After S  & 0.883 & 0.642 & 0.083 \\
&                            & After NS & 0.607 & 0.662 & 0.008 \\
\midrule

\multirow{5}{*}{S$\rightarrow$H}
& --                         & WM       & 0.997 & 0.993 & 0.044 \\
& \multirow{2}{*}{NS$\rightarrow$S}
                             & After NS & 0.659 & 0.988 & 0.009 \\
&                            & After S  & 0.615 & 0.617 & 0.148 \\
& \multirow{2}{*}{S$\rightarrow$NS}
                             & After S  & 0.897 & 0.581 & 0.136 \\
&                            & After NS & 0.611 & 0.594 & 0.009 \\
\bottomrule
\end{tabular}
\end{table}

The results in Table \ref{tab:phase3_sequential_attacks_summary_compact} support the general findings of UnMarker \cite{kassis2025unmarker}. The semantic attack is highly effective against StegaStamp. Additionally it also consistently reduces the BA of HiDDeN. This is a notable finding indicating the partial effectiveness of the UnMarker semantic attack on a non-semantic watermark. Meanwhile, the non-semantic attack has only a limited effect on StegaStamp bit accuracy.
This suggests that, among the two individual attacks evaluated, the semantic attack has the broader effect against dual-watermarked images.

Results from Phase 3 show that dual-watermarking schemes can be  attacked by sequential UnMarker attacks successfully, substantially reducing both watermark signals as shown in Table \ref{tab:phase3_sequential_attacks_summary_compact}. However, in some images, traces of one or both watermarks remain. This should be evaluated further in future studies. The experiments show that when two types of watermarks are present, a single attack does not consistently reduce both watermarks; one remains strongly decodable. Therefore, aligning with WAVES \cite{an2024waves}, a second attack is necessary to substantially weaken both watermark signals.

Both attack orders reduce the BA of HiDDeN and StegaStamp, although their relative effectiveness varies across embedding orders and decoders. For HiDDeNStega, one attack order produces a lower final HiDDeN BA, while the reverse order produces a lower StegaStamp BA. For StegaHiDDeN, applying the semantic attack followed by the non-semantic attack performs slightly better overall. Therefore, attack order has a limited and configuration-dependent effect.

The semantic attack causes more distortion but affects both decoders, whereas the non-semantic attack primarily affects HiDDeN. The results therefore support the semantic attack as the broader individual attack. Success against dual-watermarking schemes strengthens the case for UnMarker and other adversarial optimization attack schemes that operate without prior knowledge of the watermarking schemes.

Based on the results from this study, the UnMarker attacks act as complementary frequency disruptors. UnMarker formulates them to disrupt non-semantic and semantic watermark characteristics separately. HiDDeN and StegaStamp are associated with different regions of the frequency spectrum. The SSNSW schemes ensure that neither individual UnMarker attack consistently reduces both watermarks close to their respective original-image baselines. This increases robustness against watermark removal by motivating the use of two complementary attacks for complete watermark removal.

WAVES \cite{an2024waves} argues that quality cost must be accounted for when robustness is increased. The robustness added should be compared with the corresponding loss in visual quality. In this study, combining HiDDeN and StegaStamp increases robustness against the tested attacks because two complementary attacks are required to substantially reduce both watermarks. The Phase 2 results also show that the second watermark can be added with limited interference and perceptual cost. Therefore, this study is aligned with the WAVES quality--robustness principle.

\subsection{Perceptual Quality and LPIPS}
Across the experiments, LPIPS indicates limited perceptual change. However, qualitative inspection shows that it does not always reflect visually noticeable distortion.

In Phase-1, the CelebA dataset has slightly higher LPIPS than COCO, suggesting that more distortions occur in images with facial features. This is in line with the findings from UnMarker \cite{kassis2025unmarker} that the non-semantic attack creates high-frequency disruptions that can distort watermarks like HiDDeN while visual fidelity is controlled. For StegaStamp under the semantic attack, the quality inspection shows a different finding. The facial image in Fig.~\ref{fig:base_stega_low} shows considerable distortion in both the watermarked and watermark-removal images.  In contrast, the object image in Fig.~\ref{fig:base_stega_high} shows substantially less visible distortion despite a similar LPIPS value. These observations suggest that the relationship between LPIPS and visible distortion is content-dependent, with a larger discrepancy in the selected facial images.

The Phase 2 results further show that the perceptual difference is not
symmetric between the two watermarking schemes. HiDDeN mainly introduces a blurring effect, while StegaStamp creates more noticeable changes around image structures and edges. Despite this blurring effect, the overall content of the selected image remains largely unchanged after HiDDeN embedding. StegaStamp has a much larger LPIPS, although the distortion is less noticeable in the selected COCO example. This is consistent with StegaStamp's objective of maintaining decodability after physical transformations such as printing \cite{tancik2020stegastamp}. For the StegaHiDDeN pipeline, the CelebA example in Fig.~\ref{fig:p2_sh_step2} shows a similar finding. Despite the relatively low LPIPS values, the impact on the facial image is noticeable. StegaStamp does create visual artifacts, while HiDDeN only creates a blurring effect. This supports the Phase 1
observation that the selected facial images show a larger discrepancy
between LPIPS and qualitative visual quality.

Similar to semantic watermarking, the semantic attack also produces
higher visual distortion than the non-semantic attack despite
relatively low LPIPS values. LPIPS values are compared with
qualitative inspection using Figs.~\ref{fig:p3_hs_attack_pipelines}
and \ref{fig:p3_sh_attack_pipelines} that represent attacks using both
sequences. CelebA images are used in these examples to further examine
the facial-image behaviour observed in the earlier phases.
Interestingly, the semantic attack does not produce artifacts similar
to the semantic watermarking scheme StegaStamp when LPIPS is low. It is observed that the comparatively high LPIPS significantly distorts the facial image. In these selected attack examples, LPIPS corresponds more closely with the visible distortion than in the watermarking examples. Nevertheless, an LPIPS value of approximately $0.15$ still corresponds to considerable distortion in the selected facial image.

The attacks were configured to produce LPIPS values comparable to
those introduced by the watermarking schemes to keep the balance
between attack and defense. Therefore, for semantic attacks, some higher visual distortion was allowed in some cases, as StegaStamp values tend to be very high for perfect BA scores. This is in line with the objective of StegaStamp \cite{tancik2020stegastamp}. Overall, LPIPS remains useful for comparing perceptual change across
the experiments, but the qualitative examples show that it should not
be used alone to assess visual quality. In particular, the selected
facial images show larger discrepancies between LPIPS and visually
noticeable distortion than the selected object images.

\subsection{Limitations, gaps and future works}
The study is limited to evaluating only two watermarking schemes for SSNSW schemes due to the computational resources required for the experiments. HiDDen and StegaStamp were selected as the non-semantic and semantic watermarking schemes, respectively. However, the success of other semantic watermarking schemes cannot be concluded from the results of StegaStamp alone. Future studies should therefore evaluate SSNSW using other watermarking methods, such as Stable Signature and Tree-Ring Watermarks. Furthermore, the attack sequence in this study is limited to UnMarker attacks. Therefore, the robustness conclusions are limited to the evaluated UnMarker setting, and future work should test SSNSW against other watermark-removal and evasion attacks.

The second limitation concerns perceptual evaluation. As discussed, LPIPS does not always correspond to visual quality in facial-image examples. LPIPS provides a consistent metric of perceptual change. However, it should not be used alone to evaluate visual fidelity. Future studies should incorporate human-grounded perceptual evaluation alongside LPIPS, aligning with WAVES benchmarking. Alternatively, methods such as R-LPIPS \cite{ghazanfari2023r} can be investigated to determine whether they capture distortions better in face images.

The computational cost also limited the experiments to 50 images from two datasets. Each semantic attack on the 50 images required approximately three to four hours on the FAU HPC system. While the paired experimental design allows trends to be observed within the evaluated sample, the dataset size limits broader analysis of the substantial per-image variation. In particular, some images retain watermark traces with BA values above $0.70$ after attack. Larger-scale experiments could investigate which image characteristics, objects, or visual features are associated with these residual watermark traces. They could also help determine which image characteristics are associated with discrepancies between LPIPS and qualitative visual quality.

Finally, future watermarking schemes could investigate embedding information across multiple complementary frequency ranges to increase the difficulty of watermark removal while maintaining low perceptual cost. The SSNSW experiments show that combining two watermarks associated with different frequency characteristics can require two attacks for effective removal. Future work should therefore investigate whether broader spectral regions can support additional watermark representations while maintaining visual quality. The residual traces observed after sequential attacks also motivate a more detailed spectral analysis to determine where surviving watermark information remains and whether these components can be exploited to design more robust multi-layer watermarking schemes.

%% file: sections/06_conclusions.tex
\section{Conclusions}\label{chap:conclusion}

The application of SSNSW schemes increases robustness against single
UnMarker attacks. When HiDDeN and StegaStamp are embedded sequentially,
regardless of order, both watermarks remain decodable, and the
weakening of the initially embedded watermark after embedding the
second one is minimal. Embedding order therefore has only a limited effect, with StegaStamp followed by HiDDeN providing slightly better post-embedding decodability for the evaluated pair.

More importantly, neither UnMarker attack can reduce both watermark BA
scores close to their respective baselines. The semantic attack
offers more disruption to both watermarks, but not enough to remove
major traces of the non-semantic scheme. Sequential attacks are
therefore required to substantially reduce both watermarks, while
attack order has a limited and configuration-dependent effect.

SSNSW therefore raises the effort required for watermark removal but is
not a complete solution to adversarial attacks. The study also finds
that LPIPS, despite remaining low, can correspond to clear visible
distortions, especially when facial images are considered. Future work
should be more multi-layered to increase robustness, evaluate
additional watermarking and attack schemes, and use better visual
metrics alongside larger-scale experiments.